\documentclass[11pt]{article}

\usepackage{amsmath,amsthm,amssymb,cite,enumerate} 

\usepackage[usenames]{color}
\usepackage{bm}
\usepackage{authblk} 

\def \BEA { \begin{eqnarray}}
\def \EEA {\end{eqnarray}}
\def \BE {\begin{equation}}
\def \EE {\end{equation}}

\newcommand{\bL}{\mbox{\boldmath$\rho$}}

\newcommand{\LL}{\rho}
\newcommand{\NN}{N}
\newcommand{\bb}{\mbox{\boldmath$b$}}
\newcommand{\bI}{\mbox{\boldmath$I$}}

\def \WDS #1 {\mbox{$\Phi_{#1}^{S}$}}
\def \WDA #1 {\mbox{$\Phi_{#1}^{A}$}}
\def \WD #1 {\mbox{$\Phi_{#1}$}}

\def \mi {\stackrel{i}{m}}
\def \mj {\stackrel{j}{m}}
\def \mk {\stackrel{k}{m}}
\def \mr {\stackrel{r}{m}}
\def \ms {\stackrel{s}{m}}
\def \mz {\stackrel{z}{m}}
\def \mq {\stackrel{q}{m}}
\def \mo {\stackrel{o}{m}}
\def \mD {\stackrel{2}{m}}
\def \mT {\stackrel{3}{m}}
\def \mC {\stackrel{4}{m}}

\def \mio #1 {\mi_{#1}\ ^{  \! \! \! \! 0}} 
\def \mjo #1 {\mj_{#1}\ ^{  \! \! \! \! 0}} 
\def \mko #1 {\mk_{#1}\ ^{  \! \! \! \! 0}} 
\def \mro #1 {\mr_{#1}\ ^{  \! \! \! \! 0}} 
\def \mso #1 {\ms_{#1}\ ^{  \! \! \! \! 0}} 
\def \mpo #1 {\mp_{#1}\ ^{  \! \! \! \! 0}} 
\def \mzo #1 {\mz_{#1}\ ^{  \! \! \! \! 0}} 
\def \mqo #1 {\mq_{#1}\ ^{  \! \! \! \! 0}} 
\def \moo #1 {\mo_{#1}\ ^{  \! \! \! \! 0}} 
\def \mDo #1 {\mD_{#1}\ ^{  \! \! \! \! 0}} 
\def \mTo #1 {\mT_{#1}\ ^{  \! \! \! \! 0}} 
\def \mCo #1 {\mC_{#1}\ ^{  \! \! \! \! 0}} 

\def \miJ #1 {\mi_{#1}\ ^{  \! \! \! \! (1)}} 
\def \mjJ #1 {\mj_{#1}\ ^{  \! \! \! \! (1)}} 
\def \mkJ #1 {\mk_{#1}\ ^{  \! \! \! \! (1)}} 
\def \mrJ #1 {\mr_{#1}\ ^{  \! \! \! \! (1)}}

\def \bl {\mbox{\boldmath{$\ell$}}}

\def \bn {\mbox{\boldmath{$n$}}}

\def \hbm #1 {\mbox{\boldmath{$\hat m^{(#1)}$}}}
\def \bm {\mbox{\boldmath{$m$}}}

\def \Ms {\stackrel{s}{M}}

\newcommand{\be}{\begin{equation}}
\newcommand{\ee}{\end{equation}}
\newcommand{\beqn}{\begin{eqnarray}}
\newcommand{\eeqn}{\end{eqnarray}}
\newcommand{\pa}{\partial}
\newcommand{\ba}{\begin{array}}
\newcommand{\ea}{\end{array}}

\def \BEAH {\begin{eqnarray*}}
\def \EEAH {\end{eqnarray*}}
\def \BEA {\begin{eqnarray}}
\def \EEA {\end{eqnarray}}
\def \BDM {\begin{displaymath}}
\def \EDM {\end{displaymath}}

\def \Ms {\stackrel{s}{M}}

\def \T {\bigtriangleup  }

\newcommand{\M}[3] {{\stackrel{#1}{M}}_{{#2}{#3}}}

\newtheorem{proposition}{Proposition}

\begin{document}

\title{On higher dimensional Einstein spacetimes with a non-degenerate double Weyl aligned null direction}

\author[1,2]{Marcello Ortaggio\thanks{ortaggio(at)math(dot)cas(dot)cz}}
\author[1]{Vojt\v ech Pravda\thanks{pravda@math.cas.cz}}
\author[1]{Alena Pravdov\' a\thanks{pravdova@math.cas.cz}}

\affil[1]{Institute of Mathematics of the Czech Academy of Sciences, \newline \v Zitn\' a 25, 115 67 Prague 1, Czech Republic}
\affil[2]{Instituto de Ciencias F\'{\i}sicas y Matem\'aticas, Universidad Austral de Chile, \newline Edificio Emilio Pugin, cuarto piso, Campus Isla Teja, Valdivia, Chile}

\date{\today}

\maketitle

\begin{abstract}

We prove that higher dimensional Einstein spacetimes which possess a geodesic, 
non-degenerate double Weyl aligned null direction (WAND) $\bl$ must additionally 
possess a second double WAND (thus being of type D) if either:  
(a) {the Weyl tensor obeys $C_{abc[d}\ell_{e]}\ell^c=0$ ($\Leftrightarrow\Phi_{ij}=0$, i.e.,} the Weyl type is II(abd)); (b)  $\bl$ is twistfree.  
Some comments about an extension of the Goldberg-Sachs theorem to six dimensions 
are also made.

\end{abstract}



\section{Introduction and summary of main results}

\label{sec_intro}

The Petrov classification has been extended to higher dimensions {using the notions of null alignment and boost weight} \cite{Coleyetal04}  (cf. also the review \cite{OrtPraPra13rev}). {It provides one with a useful tool to classify and construct exact solutions, as well as to understand geometric properties of spacetimes (see \cite{OrtPraPra13rev} for examples of certain applications and for references).} {A class of spacetimes of particular interest consists of those of Weyl type II}. 
In any dimension $n\ge4$, the {Weyl} type II condition of \cite{Coleyetal04} can be expressed as \cite{Ortaggio09}
\be
	\ell_{[e}C_{a]b[cd}\ell_{f]}\ell^b=0 ,
	\label{mWAND}
\ee
where $\bl$ is a null vector field that defines a {\em multiple Weyl aligned null direction} (mWAND). 
{Equivalently, the Weyl type II can be defined by the existence of a null frame in which all positive boost weight (b.w.) components of the Weyl tensor vanish \cite{Coleyetal04,OrtPraPra13rev}. {In four dimensions this is equivalent to the Petrov type II, for which the highest b.w. components (i.e., b.w. 0)} are represented by the complex {scalar} $\Psi_2$. {However, in higher dimension $n$} the Weyl tensor admits $\frac{1}{12} (n-3) (n-2)  [n(n-3)+8 ] $  real b.w. 0 components. Thus the number of such components  increases rapidly with the dimension of the spacetime and, not surprisingly, new {subtypes} of type II, not occurring in four dimensions, appear.\footnote{{The definitions of the Weyl subtypes, such as D(d) and II(abd), can be found in \cite{Coleyetal04,Ortaggio09,OrtPraPra13rev}. The notation and various symbols to be used in this paper are explained at the end of this section (see also \cite{Coleyetal04,Pravdaetal04,PraPraOrt07,OrtPraPra13rev} for more details).}}
} {One {of these is the subtype} II(abd) {(considered also  in this paper)}, which can be defined by the existence of a  null vector field $\bl$ obeying
\be
 C_{abc[d}\ell_{e]}\ell^c=0 ,
\ee
{(which is stronger than \eqref{mWAND})} or equivalently by the vanishing of the part of b.w. 0 Weyl components described by the matrix $\Phi_{ij}$ (introduced in \eqref{def_Phi}).} 

While for $n=4$ large classes of type II Einstein spacetimes are known \cite{Stephanibook}, there appear to exist various restrictions on solutions of genuine type II in more than four dimensions. First, if an Einstein spacetime admits a non-geodesic mWAND, it is necessarily of type D  \cite{DurRea09} (more precisely, of type D(d), and there exists an infinity of mWANDs, including a geodesic one). Additionally, if $\bl$ is twistfree and non-degenerate (as defined in \eqref{detL} below) and $\Phi_{ij}\neq0$ (i.e., the type is {\em not} II(abd)), then again the type is necessarily D (more precisely type D(bd), see Proposition~4.1 of \cite{OrtPraPra13} and cf. also \cite{OrtPraPra09b}; this case includes, e.g., Robinson-Trautman solutions with $\mu\neq0$ \cite{PodOrt06}). In five dimensions, the same conclusion remains true even if one drops the twistfree assumption \cite{deFGodRea15} (while $\Phi_{ij}\neq0$ follows automatically for type II with $n=5$ \cite{PraPraOrt07} and need not be assumed). In six dimensions, a similar result has been obtained in \cite{Ortaggio17} in the Ricci-flat case (again without assuming the twistfree condition, but with an additional assumption on the asymptotic fall-off of the Weyl tensor -- which implies $\Phi_{ij}\neq0$  and which is automatically satisfied in five dimensions). Restrictions in the $n=5$ degenerate case have been obtained in \cite{Wylleman15}.

The purpose of the present contribution is to present some new results which hold in arbitrary higher dimensions for the case of a non-degenerate mWAND. First, we will prove the following 
\begin{proposition}
\label{prop_Phi=0} 
An $n>5$ dimensional Einstein spacetime of type II(abd) 
with a geodesic mWAND $\bl$ such that $\det \bL\neq0$ must be 
in fact of type D(abd). In a frame such that {$L_{i1}=0$}, the null vector $\bn$ is another mWAND.
\end{proposition}
Examples of such spacetimes with $n>6$ can be found in appendix~F of \cite{OrtPraPra13} (in particular, in appendix~F.4 in the twisting case). Other non-twisting examples (for $n\ge6$) are given by the Robinson-Trautman solutions with $\mu=0$ \cite{PodOrt06}.

As a further restriction on type~II spacetimes, we will additionally prove
\begin{proposition}
	\label{prop_nontwist_typeD}
	An $n>4$ dimensional Einstein spacetime that admits a non-degenerate, 
	non-twisting mWAND $\bl$ is necessarily of type D(bd) and, 
	in a frame such that {$L_{i1}=0$}, the null vector $\bn$ is another mWAND. 
	If $\Phi_{ij}\neq0$, $\bl$ must be shearfree \cite{OrtPraPra13} and $\bl$ 
	and $\bn$ are the only WANDs. If $\Phi_{ij}=0$, the type further specializes 
	to D(abd) and, for $n>6$, the spacetime may admit a continuous infinity of mWANDs. 
\end{proposition}
This is a refinement of Proposition~4.1 of \cite{OrtPraPra13}. More details are discussed in section~\ref{subsec_proof2}  (including the conditions for having an infinite number of mWANDs in the $\Phi_{ij}=0$ case). We observe that if the non-degeneracy assumption is dropped, examples of Einstein spacetimes of genuine type II can be easily constructed taking direct products or Brinkmann warps of known lower-dimensional type II solutions \cite{OrtPraPra11,OrtPraPra13rev}.  

As a by-product of the main calculations, we will also show that 
\begin{proposition}
 \label{prop_small} 
 In an $n>4$ dimensional Einstein spacetime of type III or N, 
the mWAND is necessarily  degenerate (i.e., $\det \bL=0$).
\end{proposition}
This result was already known for the type N \cite{Pravdaetal04}. For the type III it was known in the following cases: (i) for $n=5,6$ \cite{Pravdaetal04,Kubicek_thesis}; (ii) for a non-twisting mWAND (for any $n>4$) \cite{Pravdaetal04}; (iii) for a twisting mWAND with $n>6$ under certain additional ``genericity''  assumptions on the Weyl tensor \cite{Pravdaetal04}. (For (i)--(iii) and for the type N, the optical matrix must in fact have rank 2 \cite{Pravdaetal04,Kubicek_thesis}.)

The proofs of the above propositions are presented in section~\ref{sec_proofs}. In connection with these results, in section~\ref{sec_GS6D} we study some aspects of an extension of the Goldberg-Sachs theorem to six dimensions. For the case of non-degenerate mWANDS, we arrive at some restrictions on  the  optical  matrix compatible with Einstein spacetimes of type II(abd) (section~\ref{subsec_GS_II(abd)}). Some of these results can also be extended to the general type II (section~\ref{subsec_generalII}).

\paragraph{Notation} Let us summarize the Newman-Penrose notation used in the present paper. The $n$ frame vectors $\bm_{(a)}$ consist of two null vectors $\bl\equiv\bm_{(0)}$,  $\bn\equiv\bm_{(1)}$ and $n-2$ orthonormal spacelike vectors $\bm_{(i)} $, with $a, b\ldots=0,\ldots,n-1$ while $i, j  \ldots=2,\ldots,n-1$ \cite{Coleyetal04,OrtPraPra13rev}.
The Ricci rotation coefficients are defined by \cite{Pravdaetal04}
\be
 L_{ab}=\ell_{a;b} , \qquad N_{ab}=n_{a;b}  , \qquad \M{i}{a}{b}=m^{(i)}_{a;b} ,
 \label{Ricci_rot}
\ee
and satisfy the identities $L_{0a}=N_{1a}=N_{0a}+L_{1a}=\M{i}{0}{a} + L_{ia}
 = \M{i}{1}{a}+N_{ia}=\M{i}{j}{a}+\M{j}{i}{a}=0$.

Covariant derivatives along the frame vectors are denoted as
\be
	D \equiv \ell^a \nabla_a, \qquad \T\equiv n^a \nabla_a, \qquad \delta_i \equiv m_{(i)}^{a} \nabla_a . 
 \label{covder}
\ee

For the non-zero components of a Weyl tensor of type II, we use the symbols \cite{Pravdaetal04,PraPraOrt07}
\beqn
 & & \WD{ij} = C_{0i1j} , \qquad \WDS{ij} =\WD{(ij)} , \qquad \WDA{ij} =\WD{[ij]} , \qquad \WD{ } =\WD{ii} , \label{def_Phi} \\
 & & \Psi_{i} = C_{101i}, \qquad \Psi_{ijk}= \frac{1}{2} C_{1kij}, \qquad \Psi_{ij} = \frac{1}{2} C_{1i1j} , \label{def_Psi} 
\eeqn 
which satisfy the identities $C_{01ij}=2 C_{0[i|1|j]}=2\WDA{ij} $, $2C_{0(i|1|j)}=2\WDS{ij} =-C_{ikjk}$, $2C_{0101}= -C_{ijij}=2\WD{ } $, $\Psi_i=2 \Psi_{ijj}$, $\Psi_{\{ijk\}}=0$ {(curly brackets denoting cyclicity)}, $\Psi_{ijk}=-\Psi_{jik}$, $\Psi_{ij}=\Psi_{ji}$, and $\Psi_{ii}=0$. Throughout the paper, the Einstein equations $R_{ab}=\frac{R}{n}g_{ab}$ hold, so that $R_{abcd;e}=C_{abcd;e}$. It will be convenient to define the rescaled Ricci scalar
\be 
 \tilde R=\frac{R}{n(n-1)} . 
\ee

\section{Proof of Propositions~\ref{prop_Phi=0}--\ref{prop_small}}

\label{sec_proofs}

\subsection{Preliminaries: Einstein spacetimes with a non-degenerate mWAND}

Let us consider a $n$-dimensional Einstein spacetime of type II or more special ($n>4$). It follows \cite{DurRea09} that there exists a null vector field $\bl$ that satisfies \eqref{mWAND} while being geodesic and affinely parametrized, i.e., 
\be
 \ell_{a;b}\ell^b=0 .
 \label{geodesic}
\ee
The rank of the $(n-2)\times(n-2)$ {\em optical matrix} {with components}
\be
  \LL_{ij}\equiv L_{ij}=\ell_{a;b}m_{(i)}^am_{(j)}^b ,
	\label{L_def}
\ee
is thus frame-independent \cite{OrtPraPra07}, for any choice of a frame adapted to $\bl$.

In this paper we consider the {\em non-degenerate}  (or full-rank) case, i.e., from now on we assume
\be
 \det \bL\neq 0 .
 \label{detL}
\ee
We will find certain conditions under which the genuine type II cannot occur.

For convenience, we take our frame to be parallelly propagated along $\bl$ \cite{OrtPraPra07}, so that (with \eqref{geodesic}) 
\be
  L_{i0}
=0 , \qquad L_{10}=0, \qquad \M{i}{j}{0}=0, \qquad N_{i0}
=0 . 
\label{parall_transp}
\ee
Furthermore, thanks to \eqref{detL}, we can perform a null rotation about $\bl$ such that (cf. appendix~D.2.6 of \cite{OrtPraPra09} and lemma~1 of 
\cite{Durkeeetal10}) 
\be
L_{i1}
=0 .
 \label{Li1}
\ee
{Given $\bl$, this uniquely fixes} the null direction defined by $\bn$ (this choice will be useful in the following). We also define an affine parameter $r$ such that
\be
 \bl=\pa_r .
 \label{l}
\ee

With the above assumptions, the Sachs equation $D\bL=-\bL^2$ \cite{Pravdaetal04,OrtPraPra07} 
fixes the $r$-dependence of $\bL$ (thanks to the simple identity $D(\bL \bL^{-1})=0$)
\cite{NP,OrtPraPra09,OrtPraPra09b} 
\be
  \bL^{-1}=r{\bI}-\bb , 
 \label{inverse}
\ee
where $\bI$ is the identity matrix and $D\bb=0$. 
We can restrict ourselves to the case $\bb\neq0$, for the case $\bb=0$ 
reduces to the Robinson-Trautman metrics, already known to be of type D \cite{PodOrt06}. 
Note that $\bl$ is twistfree iff $b_{[ij]}=0$ \cite{OrtPraPra09b}. 
For later purposes, it is useful to observe \cite{OrtPraPra09b} that, 
for large $r$, eq.~(\ref{inverse}) gives 
\be
 \LL_{ij}=\frac{1}{r}\delta_{ij}+\frac{1}{r^2}b_{ij}+O(r^{-3}) . 
 \label{L_infty}
\ee

\subsection{Type II(abd) spacetimes: proof of Propositions~\ref{prop_Phi=0} and \ref{prop_small}}

\label{subsec_II(abd)}

Let us now focus on Einstein spacetimes of type II(abd), i.e., such that 
\be
	\Phi_{ij}=0 ,
	\label{II(abd)}
\ee	
in the frame specified above (this case is of interest only for $n>5$, since~\eqref{II(abd)} means that all b.w.~0 Weyl components vanish when $n=5$ \cite{PraPraOrt07}). 

The $r$-dependence of the Weyl tensor can be determined by using the Bianchi equations containing $D$-derivatives. Using \eqref{II(abd)} and the fact that $\bl$ is an mWAND, eq.~(B.12,\cite{Pravdaetal04}) (cf. also (7,\cite{OrtPraPra09b})) reduces to
\be
 DC_{ijkm}=-C_{ijkl}\LL_{lm}+C_{ijml}\LL_{lk} \label{DCijkm} ,
\ee
which, using \eqref{inverse}, implies $D(C_{ijkm}\LL^{-1}_{kl}\LL^{-1}_{ms})=0$. By integration one gets 
\be
 C_{ijkm}=\LL_{lk}\LL_{sm} c_{ijls} .
 \label{Cijkl}
\ee
The integration functions $c_{ijls}(=-c_{jils}=c_{lsij})$ do not depend on $r$ and, thanks to $C_{ijkj}=0$ (which follows from \eqref{II(abd)}) and \eqref{Cijkl} with \eqref{L_infty} (cf. also (27) and (22) of \cite{PraPraOrt07}), satisfy  
\beqn
 & & c_{ikjk}=0 , \label{C2} \\
 & & c_{ijkl}b_{(lj)}=0 , \qquad c_{ijkl}b_{[lj]}=0 . \label{pi3}
\eeqn

Next, eqs.~(B.1), (B.6), (B.9) and (B.4) of \cite{Pravdaetal04} (cf. also (16)--(18), (22) of \cite{OrtPraPra09b}) reduce to
\BEA
	& & D\Psi_i=-2\Psi_s \LL_{si}  \label{PII-B1}, \\
	& &  2D\Psi_{ijk}= -2\Psi_{ijs}\LL_{sk}-\Psi_{i}\LL_{jk}+\Psi_{j}\LL_{ik} \label{PII-B6}, \\
	& & D\Psi_{jki}=2\Psi_{[k|si}\LL_{s|j]}+\Psi_{i}\LL_{[jk]} , \label{PII-B9} \\
	& & 2D\Psi_{ij}=-2\Psi_{is}\LL_{sj} +\delta_{j}\Psi_i+\Psi_i L_{1j}+\Psi_s \Ms_{ij} . \label{PII-B4}
\EEA

Thanks to \eqref{inverse}, eqs.~\eqref{PII-B1}--\eqref{PII-B9} can be easily integrated, giving
\beqn
  & & \Psi_i=\LL_{lk}\LL_{ki}\psi_{l} , \label{B1_} \\
	& & \Psi_{ijk}=\LL_{l[i}\LL_{j]k}\psi_{l}+\LL_{lk}\hat\psi_{ijl} , \label{B6_} \\
	& & \Psi_{jki}=\LL_{nj}\LL_{mk}(\psi_{nmi}-b_{[nm]}\psi_{l}\LL_{li}) , \label{B9_}
\eeqn 
where $\psi_{l}$, $\hat\psi_{ijl}$ and $\psi_{nmi}$ are $r$-independent quantities. Comparing \eqref{B6_} and \eqref{B9_} (e.g., using \eqref{L_infty}) immediately gives $\hat\psi_{ijl}=0$. Using this, the trace of \eqref{B6_} with \eqref{B1_} additionally yields (for $n>4$) $\psi_{l}=0$, and therefore 
\be 
	\Psi_i=0=\Psi_{ijk} .
\ee

Thanks to these, eq.~\eqref{PII-B4} gives
\be
 \Psi_{ij}=\LL_{kj}\psi_{ik} ,
 \label{Psi_ij}
\ee
with $\psi_{[ik]}=0=\psi_{ii}$ and $D\psi_{ik}=0$. 

In order to proceed, we will also need (B.13,\cite{Pravdaetal04}) (cf. also (B11,\cite{Ortaggio17})), which here becomes
\beqn
  & & -\T C_{ijkm}= 2\Psi_{im} \LL_{jk} +  4\Psi_{[j|k} \LL_{|i]m} 
	-2\Psi_{jm} \LL_{ik}+2C_{ij[k|s} \Ms_{|m]1} +  2C_{[i|skm} \Ms_{|j]1}+2C_{ij[k|s} \NN_{s|m]} .
 \label{B13}
\eeqn 
The $r$-dependence of the last term can be fixed using the Ricci identity 
(11j,\cite{OrtPraPra07}), which with \eqref{parall_transp} and \eqref{II(abd)} 
reduces to $D\NN_{ij}=-\NN_{ik}\LL_{kj}-\tilde R\delta_{ij}$. 
This gives
\be
 \NN_{ij}=\LL_{kj}\left(-\frac{1}{2}\tilde Rr^2\delta_{ik}+\tilde Rrb_{ik}+n_{ik}\right) ,
 \label{Nij_}
\ee
where $Dn_{ik}=0$. The trace of \eqref{B13} (with \eqref{inverse}, \eqref{Cijkl}, \eqref{Psi_ij}, \eqref{Nij_} and \eqref{C2}) leads to $\psi_{ik}=0$ (thanks to $n>4$) and thus
\be
 \Psi_{ij}=0 .
\ee

We have thus proven that all Weyl components of negative b.w. vanish identically, i.e., $\bn$ is also a mWAND. Therefore we have proven Proposition~\ref{prop_Phi=0}. In passing, the above calculations also prove Proposition~\ref{prop_small}.

\subsection{Type II spacetimes with a twistfree mWAND: proof of Proposition~\ref{prop_nontwist_typeD}}

\label{subsec_proof2}

Using Proposition~\ref{prop_Phi=0}, as a refinement of Proposition~4.1 of \cite{OrtPraPra13} one additionally obtains Proposition~\ref{prop_nontwist_typeD}. 

 Let us further observe here that, in that context, a {\em shearfree} $\bl$ 
defines Robinson-Trautman spacetimes \cite{PodOrt06}, for which the type is 
D(bd) or more special and $\bl$ and $\bn$ define two mWANDs (this follows from 
\cite{PodOrt06}, cf. also \cite{OrtPraPra13,OrtPodZof15}; $\bn$ is fixed by the $L_{i1}=0$ condition). The fact that for $\Phi_{ij}\neq0$, there are no additional
 (m)WANDs follows from footnote~15 of \cite{OrtPodZof15} -- where it is also 
observed that, instead, a Weyl tensor of type D(abd) can admit an infinity 
of mWANDs (this happens when the equation $C_{ijkl}z_l=0$ admits a solution 
$z_l\neq0$, cf. appendix~F.4 of \cite{OrtPraPra13} for an example; however, 
in this case a non-zero $C_{ijkl}$ requires $n>6$ \footnote{When $n=6$, 
the condition $C_{ijkl}z_l=0$ with $C_{ikjk}=0$ implies $C_{ijkl}=0$ 
(this can be seen explicitly, e.g., using eqs.~\eqref{W}--\eqref{Y} 
given below, if one takes a frame such that $\bm_{(2)} $ is parallel to $z_i$).}).

\section{Partial extension of the Goldberg-Sachs theorem to six dimensions}
\label{sec_GS6D}

Various results partially extending the Goldberg-Sachs theorem to higher dimensions have been obtained in recent years  \cite{Pravdaetal04,PraPraOrt07,OrtPraPra09,DurRea09,OrtPraPra09b,Ortaggioetal12,OrtPraPra13,Kubicek_thesis}. Here we point out some additional restrictions which apply to the spacetimes of section~\ref{subsec_II(abd)} (in section~\ref{subsec_GS_II(abd)}) as well as to general type~II Einstein spacetimes (in section~\ref{subsec_generalII}), at least when $n=6$.

\subsection{Type II(abd) spacetimes}

\label{subsec_GS_II(abd)}

Let us consider type~II Einstein spacetimes and further assume \eqref{detL} and \eqref{II(abd)}. By Proposition~\ref{prop_Phi=0}, these spacetimes are in fact of type D(abd).

In addition to \eqref{C2} and \eqref{pi3}, the cyclicity $C_{i\{jkl\}}=0$ (or (21,\cite{PraPraOrt07})) further gives 
\be
 c_{i\{jkl\}}=0 , \qquad c_{is\{jk|}b_{s|m\}}=0 . 
 \label{cyclicity}
\ee

In any dimension $n>5$, the algebraic equations \eqref{C2}, \eqref{pi3} 
and \eqref{cyclicity} can be used to constrain the permitted forms 
of $b_{ij}$ and thus of the optical matrix $\LL_{ij}$. For the sake 
of definiteness, let us discuss explicitly here only the case of 
{\em six dimensions}. Before starting, let us observe that, in this case, 
a Weyl tensor of type D(abd) admits {\em precisely two} mWANDs 
$\bl$ and $\bn$,\footnote{This applies to any type D(abd) in six 
dimensions (i.e., even without assuming \eqref{detL}), cf. the comments 
in section~\ref{subsec_proof2}.} which must therefore \cite{DurRea09} 
be both {\em geodesics}. In an adapted frame, eqs.~(B.7,\cite{Pravdaetal04}) 
and (B.9,\cite{Pravdaetal04}) reduce to 
$C_{ijkl}L_{i1}=0=C_{ijkl}N_{i0}$,
which gives (since $n=6$) 
$L_{i1}=0=N_{i0}$
(in agreement with Proposition~\ref{prop_Phi=0}).

Now, when $n=6$, one can take as the only (ten) 
independent components of $c_{ijkl}$ 
\beqn
 & & W_{23}=c_{2323} , \qquad W_{25}=c_{2525} , \label{W} \\
 & & Z_{23}=c_{2434} , \qquad Z_{24}=c_{2343} , \qquad Z_{25}=c_{2353} , \qquad Z_{34}=c_{3242} , \qquad Z_{35}=c_{3252} , \qquad Z_{45}=c_{4252} , \label{Z} \\
 & & Y_{23}=c_{2345} , \qquad Y_{24}=c_{2453} ,	\label{Y} 
\eeqn
while all the remaining components can be determined using \eqref{C2} and the first of \eqref{cyclicity}.\footnote{Namely: $c_{2535}=-Z_{23}$,
	$c_{2545}=-Z_{24}$, $c_{2454}=-Z_{25}$, $c_{3545}=-Z_{34}$,
$c_{3454}=-Z_{35}$, $c_{3435}=-Z_{45}$,
$c_{3535}=-W_{23}-W_{25}$, $c_{3434}=W_{25}$, $c_{4545}=W_{23}$,
$c_{2534}=-Y_{24}-Y_{23}$.}

Let us perform an $r$-independent spin to align the frame vectors $\bm_{(i)} $ 
to an eigenframe of $b_{(ij)}$ and denote by $b_i$ its eigenvalues.
From the first of \eqref{pi3} we obtain
\beqn
 & & (b_3-b_4)W_{23}+(b_5-b_4)W_{25}=0, \qquad (b_2-b_5)W_{23}+(b_4-b_5)W_{25}=0, \label{eq_W23} \\
 & & (b_5-b_2)W_{23}+(b_3-b_2)W_{25}=0, \qquad (b_4-b_3)W_{23}+(b_2-b_3)W_{25}=0, \\
 & & (b_4-b_5)Z_{23}=0, \qquad (b_3-b_5)Z_{24}=0, \qquad (b_3-b_4)Z_{25}=0, \\
 & & (b_2-b_5)Z_{34}=0 , \qquad (b_2-b_4)Z_{35}=0, \qquad (b_2-b_3)Z_{45}=0 ,
\eeqn
while the second of \eqref{cyclicity} (using the second of \eqref{pi3} 
to get rid of terms containing $b_{[ij]}$) gives 
\beqn
 & & (b_2-b_4)Y_{23}+(b_2-b_3)Y_{24}=0, \qquad (b_3-b_5)Y_{23}+(b_4-b_5)Y_{24}=0 , \\ 
 & & (b_2-b_4)Y_{23}+(b_5-b_4)Y_{24}=0, \qquad (b_5-b_3)Y_{23}+(b_2-b_3)Y_{24}=0 , \\
 & & (b_3-b_2)Z_{23}=0, \qquad (b_4-b_2)Z_{24}=0, \qquad (b_5-b_2)Z_{25}=0, \\
 & & (b_4-b_3)Z_{34}=0 , \qquad (b_5-b_3)Z_{35}=0, \qquad (b_5-b_4)Z_{45}=0 . \label{eq_Z45}
\eeqn

It can be immediately seen from \eqref{eq_W23}--\eqref{eq_Z45} that having (at least one) $Z_{ij}$ non-vanishing requires 
that at least two pairs of $b_i$ coincide, and the same conclusion holds also for $W_{ij}$ and $Y_{ij}$. Thus, a non-zero Weyl 
tensor is compatible only with the two following multiplicities of the eigenvalues of $b_{(ij)}$: $\{a,a,a,a\}$ or $\{a,a,b,b\}$. These two cases give, respectively:
\begin{enumerate}

	\item\label{GS_generic} $b_{(ij)}=b_0\delta_{ij}$ (i.e., all the eigenvalues of $b_{(ij)}$ coincide), which is equivalent (when \eqref{detL} holds) to the so called {\em optical constraint} \cite{OrtPraPra09,OrtPraPra13}. In this case, eqs.~\eqref{eq_W23}--\eqref{eq_Z45} are identically satisfied. One can shift the affine parameter 
	$r\mapsto r+b_0$, so that $b_{(ij)}=0$. Using $r$-independent spins, one can thus arrive at the canonical form (cf., e.g., \cite{OrtPraPra09,OrtPraPra13,OrtPraPra10,Ortaggio17})
\beqn
 b = {\rm diag}\left(\left[\begin{array}{cc} 0 & b_{23} \\ 
   -b_{23} & 0 \\
  \end {array}
 \right] ,
\left[\begin {array}{cc} 0 & b_{45} \\ 
   -b_{45} & 0 \\ \end {array}
 \right]\right) .
 \label{b_eigenf}
\eeqn
	
	For $b_{23}=0=b_{45}$, $\bl$ is expanding, twistfree and shearfree, which corresponds to the Robinson-Trauman spacetimes with $\mu=0$ \cite{PodOrt06}.

	\item\label{GS_special} $b_3=b_2\neq b_5=b_4$, in which case the Weyl tensor must obey (thanks to~\eqref{eq_W23}--\eqref{eq_Z45})
	 \be
	   W_{23}=Z_{24}=Z_{25}=Z_{34}=Z_{35}=Y_{23}=0 .
		\label{Weyl_special}
	 \ee
	 (Other possible cases trivially correspond to reordering the frame vectors -- these need not be discussed separately.) Using~\eqref{Weyl_special}, the second of \eqref{pi3} reads 
\beqn
 & & -b_{24}W_{25}-b_{35}Y_{24}+b_{34}Z_{23}+b_{25}Z_{45}=0 , \qquad b_{25}W_{25}-b_{34}Y_{24}-b_{35}Z_{23}+b_{24}Z_{45}=0 ,  \label{2.5_special_1} \\
 & & b_{34}W_{25}-b_{25}Y_{24}+b_{24}Z_{23}-b_{35}Z_{45}=0 , \qquad b_{35}W_{25}+b_{24}Y_{24}+b_{25}Z_{23}+b_{34}Z_{45}=0 .  \label{2.5_special_2}
\eeqn
A non-zero Weyl tensor requires the determinant associated to the above system to vanish, i.e., 
$b_{35}=\mp b_{24}$ and $b_{34}=\pm b_{25}$.
There are therefore two possible subcases.
	\begin{enumerate}

		\item $b_{35}=b_{24}=b_{34}=b_{25}=0$, in which case \eqref{2.5_special_1} and \eqref{2.5_special_2} are identically satisfied and 
					\beqn
 b = {\rm diag}\left(\left[\begin{array}{cc} b_2 & b_{23} \\ 
   -b_{23} & b_2 \\
  \end {array}
 \right] ,
\left[\begin {array}{cc} b_4 & b_{45} \\ 
   -b_{45} & b_4 \\ \end {array}
 \right]\right) .
\eeqn
The matrices $\bb$ and $\bL$ (recall~\eqref{inverse}) are {\em normal}. 
Redefining $r$ one can set $b_2=0$ (or $b_4=0$) or $b_2+b_4=0$. In the case of zero twist ($b_{23}=0=b_{45}$), it has been shown that this case cannot occur \cite{ReaGraTur13}.

		\item  $b_{35}=\mp b_{24}$, $b_{34}=\pm b_{25}$, $b_{24}^2+b_{25}^2\neq0$, in which case \eqref{2.5_special_1} and \eqref{2.5_special_2} further restrict the Weyl tensor, namely
		\be
			Y_{24}=\pm W_{25} , \qquad Z_{45}=\mp Z_{23} .
		\ee 
  In this case $\bb$ and $\bL$ are not normal matrices. It can be verified that the symmetric matrix
	$\bL+\bL^T$ admits two pairs of repeated eigenvalues (although it is not diagonal in the frame in use). By an $r$-independent spin in the plane (23) or
	(45), one can set $b_{24}=0$ or $b_{25}=0$, but not both (and, as above, redefining 
	$r$ one can set $b_2=0$ or $b_4=0$).
	
	\end{enumerate}

\end{enumerate}

\subsection{Comments for general type II}

\label{subsec_generalII}

Proving an extension of the Goldbers-Sachs theorem in the case of general type II (i.e., without assuming \eqref{II(abd)}) would deserve a separate investigation. However, let us make here some preliminary comments, still assuming~\eqref{detL}. The $r$-dependence of $C_{ijkl}$ is not given anymore by~\eqref{Cijkl}, nevertheless for $r\to\infty$ one still has $C_{ijkl}=c_{ijkl}r^{-2}+\ldots$ with $c_{ikjk}=0$, while $\WDS{ij} ={\tau_{ij}}r^{-4}+\ldots$, $\WDA{ij} =\pi_{ij}r^{-3}+\ldots$ (where $D\tau_{ij}=0=D\pi_{ij}$, {$\tau_{[ij]}=0=\pi_{(ij)}$}) \cite{OrtPraPra09b,OrtPra14}.\footnote{In order to arrive at such results, one typically assumes that the various quantities have a power-like behaviour at the leading (and sometimes sub-leading) order, cf. \cite{OrtPra14} for more details.} In the special case $c_{ijkl}=0$, 
it has been already shown that the optical matrix $\LL_{ij}$ must obey the optical constraint \cite{OrtPraPra09b}, i.e., \eqref{b_eigenf} holds. In the generic case with both $c_{ijkl}\neq 0$ and $\Phi_{ij}\neq 0$ (so far unexplored), we can adapt some of the arguments of section~\ref{subsec_GS_II(abd)} to obtain at least a weaker result which restricts the multiplicity 
of the eigenvalues of $b_{(ij)}$. We only sketch the line of the arguments, since the details would be similar to section~\ref{subsec_GS_II(abd)}.

In this case, the second relations of both \eqref{pi3} and \eqref{cyclicity} are replaced by $c_{ijkl}b_{[jl]}=(n-4) \pi_{ik}$ and $c_{im\{jk|} b_{m|l\}}=-2\pi_{\{jk}\delta_{l\}i}$, respectively (while the first of \eqref{pi3} and \eqref{cyclicity} are unchanged; cf. eqs.~(21,22,27,\cite{PraPraOrt07}), (10,\cite{OrtPraPra09b}) and \cite{OrtPra14}), nevertheless one can still arrive at \eqref{eq_W23}--\eqref{eq_Z45} by simple algebraic manipulations. As argued above, 
it follows that either: (i) $b_{(ij)}=b_0\delta_{ij}$, i.e., 
the optical constraint holds or (ii) $b_3=b_2\neq b_5=b_4$, in which case the Weyl tensor is constrained by $W_{23}=Z_{24}=Z_{25}=Z_{34}=Z_{35}=Y_{23}=0$ (up to relabeling the frame vectors).

In both cases, {\em $b_{(ij)}$ possesses at least two pairs of repeated eigenvalues}.

\section*{Acknowledgments}

This work has been supported by research plan RVO: 67985840. The authors acknowledge support from the
Albert Einstein Center for Gravitation and Astrophysics, Czech Science Foundation GACR 14-37086G.
 The stay of M.O. at Instituto de Ciencias F\'{\i}sicas y Matem\'aticas, Universidad Austral de Chile has been supported by CONICYT PAI ATRACCI{\'O}N DE CAPITAL HUMANO AVANZADO DEL 
 EXTRANJERO Folio 80150028.

\renewcommand{\thesection}{\Alph{section}}
\setcounter{section}{0}

\renewcommand{\theequation}{{\thesection}\arabic{equation}}


\end{document}